# Alternative implementation of simplified Brillouin optical correlation-domain reflectometry

Neisei Hayashi, Yosuke Mizuno, *Member IEEE*,
and Kentaro Nakamura, *Member IEEE*

*Precision and Intelligence Laboratory, Tokyo Institute of Technology,
4259 Nagatsuta-cho, Midori-ku, Yokohama 226-8503, Japan*

Manuscript received August xx, 2014; revised xxxxx xx, 2014; accepted xxxxx xx, 2014. Date of publication xxxxx xx, 2014; date of current version xxxxx xx, 2014. This work was partially supported by Grants-in-Aid for Young Scientists (A) (no. 25709032) and for Challenging Exploratory Research (no. 26630180) from the Japan Society for the Promotion of Science (JSPS) and by research grants from the General Sekiyu Foundation, the Iwatani Naoji Foundation, and the SCAT Foundation. N.H. acknowledges a Grant-in-Aid for JSPS Fellows (no. 25007652). Corresponding author: N. Hayashi (e-mail: hayashi@sonic.pi.titech.ac.jp).

**Abstract:** We developed an alternative configuration of simplified Brillouin optical correlation-domain reflectometry, which can overcome the drawbacks of the original configuration. This system uses, as reference light, the light that is Fresnel reflected at a partial reflection point artificially produced near an optical circulator. We show that the influence of the 0th correlation peak fixed at the partial reflection point can be suppressed by replacing the nearby fibers with other fibers having different Brillouin frequency shift values (here, multi-mode fibers are used). Finally, we demonstrate a distributed measurement for detecting a 1.46-m-long strained section with a high signal-to-noise ratio.

## 1. Introduction

During the past several decades, Brillouin scattering in optical fibers has been used as an operating principle in a number of devices and systems, one of the successful application examples being distributed strain and temperature sensors. Researchers have developed various sensing techniques, which can be roughly divided into five categories: (1,2) Brillouin optical time-domain analysis (BOTDA) [1-7] and reflectometry (BOTDR) [8-11], (3) Brillouin optical frequency-domain analysis (BOFDA) [12-15], and (4,5) Brillouin optical correlation-domain analysis (BOCDA) [16-19] and reflectometry (BOCDR) [20-26]. For BOTDA and BOTDR, the measurement range extends to several tens of kilometers; however, the spatial resolution of their basic configurations is inherently limited to ~1 m because of Brillouin linewidth broadening [27, 28], although some techniques have recently been developed to mitigate this limitation [3-7, 9-11]. BOFDA and BOCDA are, in principle, free from the linewidth broadening limitation at spatial resolutions above 1 m; to date, ~3 cm [13] and

~1.6 mm [17] resolution scales have been achieved, respectively. In standard BOFDA and BOCDA systems, however, probe light in addition to pump light needs to be injected into the fiber under test (FUT), leading to their complicated and/or expensive setup. Although linear configurations have been developed [15, 19], both the pump and the probe light beams still need to be used. These problems may be resolved by resorting to the fifth technique, i.e., BOCDR.

BOCDR [20-26] utilizes the correlation control of continuous waves (Brillouin-Stokes light and reference light) [29]. Its advantages include substantial one-end accessibility, high spatial resolution (13 mm with a silica fiber [21] and 6 mm with a tellurite fiber with a high Brillouin gain coefficient [23]), and a high sampling rate. In addition, this system is simple and inexpensive. The last advantage is offered by virtue of the fact that the basic BOCDR setup can be implemented without using relatively expensive devices such as electro-optic modulators (optical-pulse generators, single-sideband modulators, etc.) and vector network analyzers (cf. BOFDA). Recently, simplified (S-) BOCDR has been developed to further enhance the BOCDR utility [30]. This configuration uses, as reference light, Fresnel-reflected light at the FUT's open end and does not include an additional reference path used in standard implementations. Although the feasibility of distributed measurement has been experimentally demonstrated, S-BOCDR suffers from two disadvantages: (1) measurement can be performed only along half of the FUT length (distal from the open end), and (2) measurement cannot be continued when the FUT has even one breakage point.

In this work, to overcome such drawbacks, we develop an alternative configuration of S-BOCDR, where the light that is Fresnel reflected at a partial reflection point artificially produced near an optical circulator is used as a reference light. Using this system, we find that it is possible to measure another half of the FUT length (proximal to the open end), which is more convenient for practical applications. Even when the FUT has a breakage point, measurement can be conducted at least up to that point. We show that the influence of the 0th correlation peak fixed at the partial reflection point can be suppressed by replacing the nearby fibers with other fibers having different Brillouin frequency shift (BFS) values. Then, we demonstrate a distributed measurement for detecting a 1.46-m-long strained section with a high signal-to-noise (SN) ratio.

## 2. Principle

Brillouin scattering in optical fibers is caused by acoustic-optical interaction, generating backscattered Stokes light, the spectrum of which is referred to as the Brillouin gain spectrum (BGS) [31]. The central frequency of the BGS is shifted downward relative to the incident pump frequency by the amount termed the BFS, which is approximately 10.8 GHz in silica single-mode fibers (SMFs) at 1.55 μm. If strain (or temperature change) is applied to a silica SMF, the BFS shifts toward a higher frequency with +580 MHz/% [32] (or +1.18 MHz/K [33]) at 1.32 μm, corresponding to +493 MHz/% (or +1.00 MHz/K) at 1.55 μm. Thus, the BFS distribution measured along a fiber provides information regarding the applied strain or temperature distribution.

A basic BOCDR setup is depicted in Fig. 1(a) in Ref. [20]. The beat signal of Stokes and reference lights is detected as a BGS. By sinusoidal modulation of the laser frequency, correlation peaks are periodically distributed along the FUT [29]; the distance between the adjacent peaks is set to be longer than the FUT length, so that only one correlation peak exists in the FUT. Correlation peaks of any order can be located in the FUT by controlling the optical path-length difference between the pump and the reference paths. The correlation peak position can be scanned along the FUT by sweeping the frequency of the sinusoidal modulation; thus, a distributed BGS measurement can be performed. As has been discussed in detail and summarized in Ref. [30], the measurement range, spatial resolution, theoretical value of the highest resolution, range-to-resolution ratio (effective number of sensing points), and theoretical value of the highest ratio are expressed as

$$d_{\text{BOCDR}} = \frac{c}{2\,n\,f_{\text{m}}}, \tag{1}$$

$$\Delta z_{\text{BOCDR}} = \frac{c\,\Delta \nu_{\text{B}}}{2\,\pi\,n\,f_{\text{m}}\,\Delta f}, \tag{2}$$

$$\Delta z_{\text{BOCDR}}^{\min} = \frac{c}{\pi\, n\, BFS}, \tag{3}$$

$$N_{\text{BOCDR}} = \frac{\pi\, \Delta f}{\Delta \nu_B}, \tag{4}$$

$$N_{\text{BOCDR}}^{\max} = \frac{\pi\, BFS}{2\, \Delta \nu_B}, \tag{5}$$

respectively, where $c$ is the light velocity in vacuum, $n$ is the core refractive index, $f_m$ is the modulation frequency of the laser frequency, $\Delta \nu_B$ is the Brillouin linewidth, and $\Delta f$ is the modulation amplitude of the laser frequency.

Fig. 1(b) shows a setup of S-BOCDR [30]. In Fig. 1(b), the laser output is injected directly into the FUT, and the reflected light (consisting not only of the Brillouin-Stokes light but also of the reference light that is Fresnel reflected at the FUT open end) is detected. When the laser frequency is subject to sinusoidal modulation, correlation peaks are generated along the FUT. As the 0th correlation peak is fixed at a zero-optical-path-difference point, i.e., at the FUT open end, some loss is artificially applied near the open end to suppress its influence. The initial configuration of S-BOCDR has two drawbacks: (1) the measurement range is limited to half of the FUT length (distal from the open end) and (2) the measurement becomes non-feasible when the FUT has a breakage point. As was theoretically derived in Ref. [30], the measurement range, spatial resolution, theoretical value of the highest resolution, range-to-resolution ratio, and theoretical value of the highest ratio are given by

$$d_{\text{S-BOCDR}} = \frac{L}{2}, \tag{6}$$

$$\Delta z_{\text{S-BOCDR}} = \frac{\Delta \nu_B\, (L-l)}{\pi\, \Delta f}, \tag{7}$$

$$\Delta z_{\text{S-BOCDR}}^{\min} = \frac{2\, \Delta \nu_B\, (L-l)}{\pi\, BFS}, \tag{8}$$

$$N_{\text{S-BOCDR}} = \frac{\pi\, \Delta f}{2\, \Delta \nu_B}, \tag{9}$$

$$N_{\text{S-BOCDR}}^{\max} = \frac{\pi\, BFS}{4\, \Delta \nu_B}, \tag{10}$$

respectively, where $L$ is the FUT length, and $l$ is the sensing position in the FUT (see Fig. 3 in Ref. [30] for the definition of $l$).

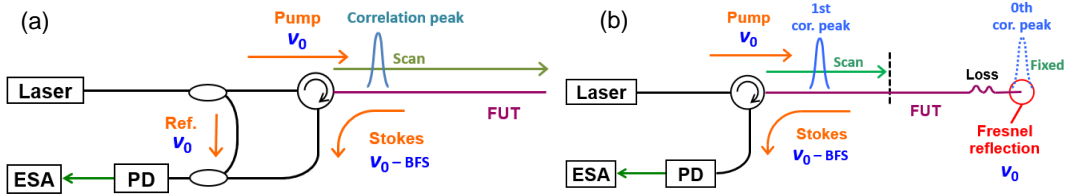

Fig. 1. Schematics of (a) basic BOCDR configuration and (b) initial S-BOCDR configuration. ESA, electrical spectrum analyzer; FUT, fiber under test; PD, photo diode.

## 3. Proposal

Schematic presentation of an alternative S-BOCDR setup is shown in Fig. 2. Similar to the case in S-BOCDR (Fig. 1(b)), the laser output here is directly injected into the FUT, and the reflected light is guided to a photodiode (PD) followed by an electrical spectrum analyzer (ESA). At the interface between the pigtail of the optical circulator and one end of the FUT, partial reflection is artificially induced, whereas the other end of the FUT is angle-cut to suppress the Fresnel reflection and is kept open. Then, the reflected light consists of the Brillouin-Stokes light and the light that is Fresnel reflected at the partial reflection point, which works as a reference light. When sinusoidal modulation is applied to the laser frequency, correlation peaks appear along the FUT. The 0th correlation peak, generally fixed at a zero-optical-path-difference point, is located at the partial reflection point in this case. As shown in Fig. 2, the influence of the 0th peak needs to be suppressed by replacing the nearby fibers with fibers that have different BFS values (unless low-temperature measurement is aimed at, lower BFS is preferable because the BFS in silica SMFs increases with increasing applied strain [32] and increasing temperature [33]). As the modulation frequency increases, the 1st correlation peak gradually approaches the 0th peak. While only the 1st peak (except the 0th peak) exists in the FUT, a distributed measurement can be properly performed. When the 1st peak has reached the FUT midpoint, the 2nd peak starts to appear at the FUT open end, resulting in the measurement error. By the same reasoning, the length of the circulator pigtail needs to be sufficiently shorter than half of the FUT length. Thus, the measurement range of the alternative S-BOCDR is limited to half of the FUT length $L$, i.e., Eq. (6) is valid; however, a major difference is that the measurement range is located proximal to the open end, which is more convenient for practical applications. The spatial resolution can be expressed by Eq. (7), provided the sensing position $l$ in the FUT is defined as shown in Fig. 3. Consequently, the theoretical value of the highest resolution, range-to-resolution ratio, and theoretical value of highest ratio of the alternative S-BOCDR are also expressed by Eqs. (8), (9), and (10), respectively. It should be noted that the range-to-resolution ratio is calculated by using the lowest resolution ($l = 0$), because the resolution dramatically depends on the sensing position $l$.

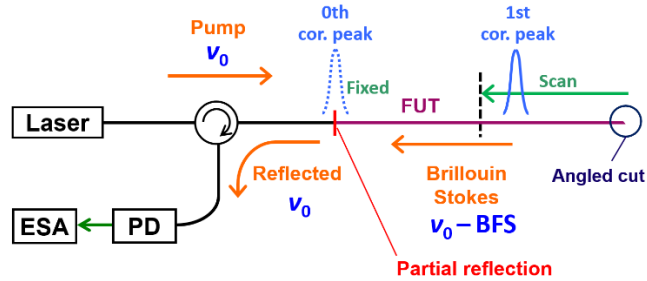

Fig. 2. Schematic of alternative S-BOCDR.

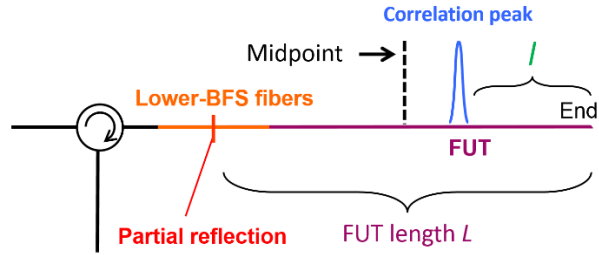

Fig. 3. Definitions of $L$ and $l$.

## 4. Experiments

Schematic of the alternative S-BOCDR setup used in the experiment is shown in Fig. 4. The pump light was amplified up to ~30 dBm using an erbium-doped fiber amplifier (EDFA). The sinusoidal modulation of the laser frequency was performed by directly modulating the laser driving current.

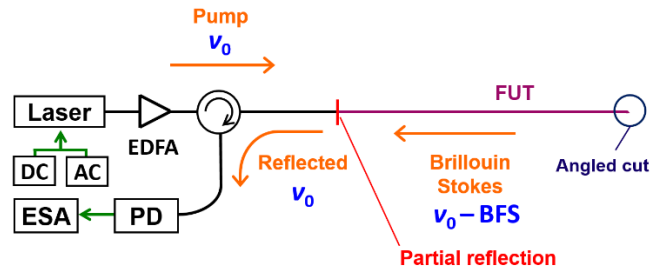

Fig. 4. Experimental setup of alternative S-BOCDR.

First, a pilot distributed measurement was performed without suppressing the influence of the 0th correlation peak. The FUT structure is shown in Fig. 5. We employed a silica SMF as an FUT, which had a numerical aperture (NA) of 0.13, a core refractive index of ~1.46, a core diameter of 9 µm, a cladding diameter of 125 µm, and a propagation loss of ~0.5 dB/km at 1.55 µm. Partial reflection was induced by forming a small gap (reflectivity ~0.05%). The FUT length was 18.00 m, leading to the measurement range of 9.00 m. In this study, to clearly show a distributed measurability with a high SN ratio, we swept the modulation frequency $f_m$ from 5.7038 MHz to 11.408 MHz and fixed the modulation amplitude $\Delta f$ at 0.50 GHz, corresponding to the spatial resolution of ~170 mm at $l$ = 0 m and ~340 mm at $l$ = 9.00 m. Using two translation stages, we applied different strains of <0.7% to a 1.46-m-long section of the FUT. The sampling rate at a single sensing point was 3.3 Hz (limited by the data acquisition from the ESA). There were 45 sensing points, and the measurement lasted for ~14 s. The operating temperature was maintained at 28 °C.

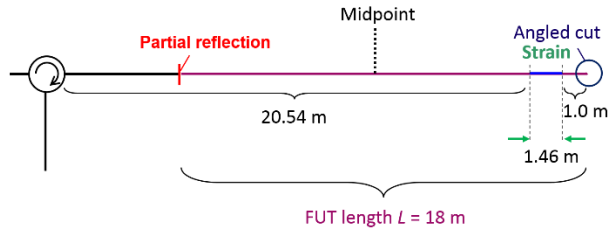

Fig. 5. Structure of the FUT without 0th peak suppression.

The measured BGS distribution with an applied strain of 0.50% is shown in Fig. 6(a), and the measured BGSs at $l$ = 8.00 m (no strain) and at $l$ = 1.40 m (strained) are shown in Fig. 6(b). Irrespective of the sensing position, the peak frequency of the BGS was 10.86 GHz, corresponding to the BFS of a non-strained silica SMF. Thus, the strained section was not detected, because the measured BGS was constantly overlapped by the BGS of the light that was Brillouin scattered at the 0th correlation peak. This result indicates that suppression of the 0th peak is indispensable for the correct operation of the alternative S-BOCDR.

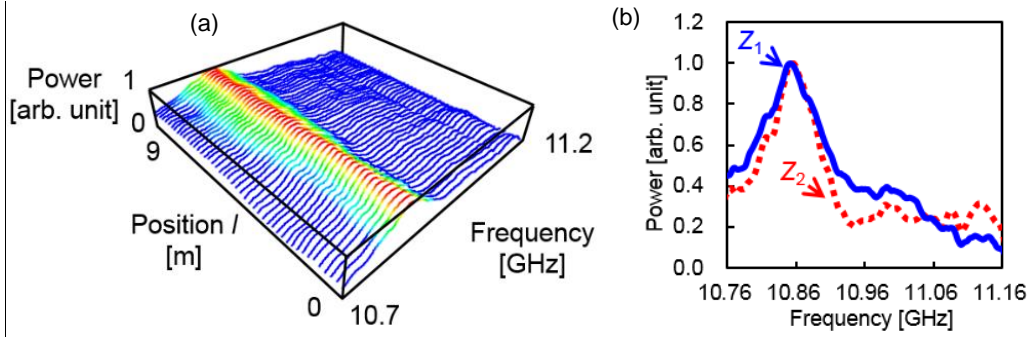

Fig. 6. (a) Measured BGS distribution without 0th peak suppression. (b) Measured BGSs at $l$ = 8.00 m ($Z_1$; no strain) and at $l$ = 1.40 m ($Z_2$; strained).

Next, the distributed measurement was performed after suppressing the influence of the 0th correlation peak. As shown in Fig. 7, two 3.00-m-long silica multi-mode fibers (MMFs) with a BFS value of ~10.5 GHz [34] were inserted at around the partial reflection point; this BFS was sufficiently lower than that of silica SMFs (corresponding to the BFS of silica SMFs at roughly –330 °C). These MMFs had an NA of 0.20, a core refractive index of ~1.46, a core diameter of 50 µm, a cladding diameter of 125 µm, and a propagation loss of ~1.0 dB/km at 1.55 µm. The two interfaces of the MMFs and SMFs were connected by adaptors. Other conditions were the same as those in the preceding experiment.

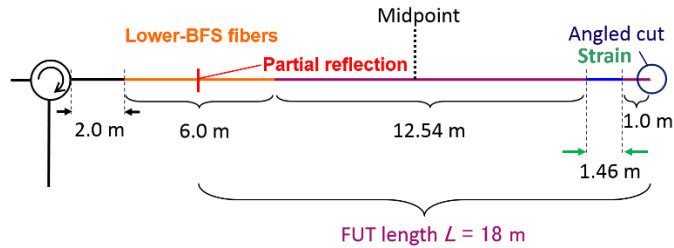

Fig. 7. Structure of the FUT with 0th peak suppression.

The measured BGS distribution with an applied strain of 0.50% is shown in Fig. 8(a), and the measured BGSs at $l$ = 8.00 m (no strain) and at $l$ = 1.40 m (strained) are shown in Fig. 8(b). A BFS upshift was clearly observed at the strained section. The measured BFS distributions for applied strains of 0.00, 0.39, 0.50, and 0.63% (Fig. 9) revealed that the 1.46-m-long strained section was successfully detected. As shown in the inset of Fig. 9, the BFS shifted to higher frequencies with increasing strain with a proportionality constant of 491 MHz/% (calculated using the BFS values at the midpoint of the strained section), which agrees well with the reported value [32]. The BFS fluctuations at the non-strained sections were approximately ±2.3 MHz (standard deviation), corresponding to the strain and temperature measurement errors of ~±0.005% and ~±2.3 °C, respectively. Thus, a distributed measurement capability of the alternative S-BOCDR was demonstrated.

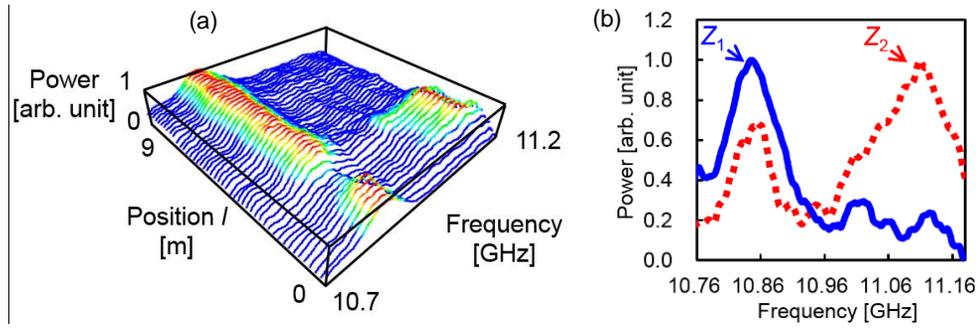

Fig. 8. (a) Measured BGS distribution with 0th peak suppression. (b) Measured BGSs at $l$ = 8.00 m ($Z_1$; no strain) and at $l$ = 1.40 m ($Z_2$; strained).

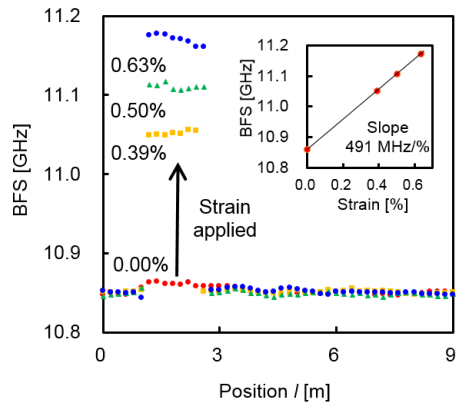

Fig. 9. Measured BFS distribution when strains of 0.00, 0.39, 0.50, and 0.63% were applied. The inset shows the BFS dependence on the applied strain.

## 5. Conclusion

We developed an alternative configuration of S-BOCDR, which uses, as reference light, the light that is Fresnel reflected at a partial reflection point artificially produced near an optical circulator. Using this system, we find that it is possible to measure half of the FUT length located proximal to the open end, which is more convenient for practical applications. Besides, measurements can be performed even when the FUT has a breakage point. We clarified that the theoretical performances of the alternative S-BOCDR can be given by expressions similar to those of the initial S-BOCDR configuration. We then experimentally demonstrated that the suppression of the influence of the 0th correlation peak fixed at the partial reflection point is crucial for successful measurement. We also demonstrated that this suppression can be achieved by replacing the fibers around the partial reflection point with other fibers having different BFS values, such as silica MMFs. Finally, we demonstrated a distributed measurement and successfully detected a 1.46-m-long strained section with a high SN ratio. We believe that, owing to its simplicity and cost efficiency, the alternative S-BOCDR will become a promising technique for practical applications of distributed Brillouin sensing.

## References

[1] T. Horiguchi and M. Tateda, "BOTDA–nondestructive measurement of single-mode optical fiber attenuation characteristics using Brillouin interaction: theory," J. Lightwave Technol. **7**, 1170–1176 (1989).
[2] T. Horiguchi, K. Shimizu, T. Kurashima, M. Tateda, and Y. Koyamada, "Development of a distributed sensing technique using Brillouin scattering," J. Lightwave Technol. **13**, 1296–1302 (1995).

[3] A. Vedadi, D. Alasia, E. Lantz, H. Maillotte, L. Thevenaz, M. Gonzalez-Herraez, and T. Sylvestre, "Brillouin optical time-domain analysis of fiber-optic parametric amplifiers," IEEE Photon. Technol. Lett. **19**, 179–181 (2007).

[4] A. W. Brown, B. G. Colpitts, and K. Brown, "Dark-pulse Brillouin optical time-domain sensor with 20-mm spatial resolution," J. Lightwave Technol. **25**, 381–386, (2007).

[5] W. Li, X. Bao, Y. Li, and L. Chen, "Differential pulse-width pair BOTDA for high spatial resolution sensing," Opt. Express **16**, 21616–21625 (2008).

[6] T. Sperber, A. Eyal, M. Tur, and L. Thevenaz, "High spatial resolution distributed sensing in optical fibers by Brillouin gain-profile tracing," Opt. Express **18**, 8671–8679 (2010).

[7] K. Y. Song and H. J. Yoon, "High-resolution Brillouin optical time domain analysis based on Brillouin dynamic grating," Opt. Lett. **35**, 52–54 (2010).

[8] T. Kurashima, T. Horiguchi, H. Izumita, and M. Tateda, "Brillouin optical-fiber time domain reflectometry," IEICE Trans. Commun. **E76-B**, 382–390 (1993).

[9] K. Shimizu, T. Horiguchi, Y. Koyamada, and T. Kurashima, "Coherent self-heterodyne Brillouin OTDR for measurement of Brillouin frequency shift distribution in optical fibers," J. Lightwave Technol. **12**, 730–736 (1994).

[10] M. N. Alahbabi, Y. T. Cho, and T. P. Newson, "150-km-range distributed temperature sensor based on coherent detection of spontaneous Brillouin backscatter and in-line Raman amplification," J. Opt. Soc. Amer. B **22**, 1321–1324 (2005).

[11] M. A. Soto, G. Bolognini, and F. D. Pasquale, "Analysis of optical pulse coding in spontaneous Brillouin-based distributed temperature sensors," Opt. Express **16**, 19097–19111 (2008).

[12] D. Garus, K. Krebber, and F. Schliep, "Distributed sensing technique based on Brillouin optical-fiber frequency-domain analysis," Opt. Lett. **21**, 1402–1404 (1996).

[13] R. Bernini, A. Minardo, and L. Zeni, "Distributed sensing at centimeter-scale spatial resolution by BOFDA: measurements and signal processing," IEEE Photon. J. **4**, 48–56 (2012).

[14] A. Minardo, R. Bernini, and L. Zeni, "Distributed temperature sensing in polymer optical fiber by BOFDA," IEEE Photon. Technol. Lett. **24**, 387–390 (2014).

[15] A. Wosniok, Y. Mizuno, K. Krebber, and K. Nakamura, "L-BOFDA: a new sensor technique for distributed Brillouin sensing," Proc. SPIE **8794**, 879431 (2013).

[16] K. Hotate and T. Hasegawa, "Measurement of Brillouin gain spectrum distribution along an optical fiber using a correlation-based technique – Proposal, experiment and simulation –," IEICE Trans. Electron. **E83-C**, 405–412 (2000).

[17] K. Y. Song, Z. He, and K. Hotate, "Distributed strain measurement with millimeter-order spatial resolution based on Brillouin optical correlation domain analysis," Opt. Lett. **31**, 2526–2528 (2006).

[18] K. Hotate and M. Tanaka, "Distributed fiber Brillouin strain sensing with 1-cm spatial resolution by correlation-based continuous-wave technique," IEEE Photon. Technol. Lett. **14**, 179–181 (2002).

[19] K. Y. Song and K. Hotate, "Brillouin optical correlation domain analysis in linear configuration," IEEE Photon. Technol. Lett. **20**, 2150–2152 (2008).

[20] Y. Mizuno, W. Zou, Z. He, and K. Hotate, "Proposal of Brillouin optical correlation-domain reflectometry (BOCDR)," Opt. Express **16**, 12148–12153 (2008).

[21] Y. Mizuno, Z. He, and K. Hotate, "One-end-access high-speed distributed strain measurement with 13-mm spatial resolution based on Brillouin optical correlation-domain reflectometry," IEEE Photon. Technol. Lett. **21**, 474–476 (2009).

[22] Y. Mizuno, W. Zou, Z. He, and K. Hotate, "Operation of Brillouin optical correlation-domain reflectometry: theoretical analysis and experimental validation," J. Lightwave Technol. **28**, 3300–3306 (2010).

[23] Y. Mizuno, Z. He, and K. Hotate, "Distributed strain measurement using a tellurite glass fiber with Brillouin optical correlation-domain reflectometry," Opt. Commun. **283**, 2438–2441 (2010).

[24] Y. Mizuno, Z. He, and K. Hotate, "Measurement range enlargement in Brillouin optical correlation-domain reflectometry based on temporal gating scheme," Opt. Express **17**, 9040–9046 (2009).

[25] Y. Mizuno, Z. He, and K. Hotate, "Measurement range enlargement in Brillouin optical correlation-domain reflectometry based on double-modulation scheme," Opt. Express **18**, 5926–5933 (2010).

[26] N. Hayashi, Y. Mizuno, and K. Nakamura, "Distributed Brillouin sensing with centimeter-order spatial resolution in polymer optical fibers," J. Lightwave Technol. (in press) [DOI: 10.1109/JLT.2014.2339361].

[27] A. Fellay, L. Thevenez, M. Facchini, M. Nikles, and P. Robert, "Distributed sensing using stimulated Brillouin scattering: Towards ultimate resolution," Tech. Dig. Opt. Fiber Sens. **16**, 324–327 (1997).

[28] H. Naruse and M. Tateda, "Trade-off between the spatial and the frequency resolutions in measuring the power spectrum of the Brillouin backscattered light in an optical fiber," Appl. Opt. **38**, 6516–6521 (1999).

[29] K. Hotate, "Application of synthesized coherence function to distributed optical sensing," Meas. Sci. Technol. **13**, 1746–1755 (2002).

[30] N. Hayashi, Y. Mizuno, and K. Nakamura, "Simplified configuration of Brillouin optical correlation-domain reflectometry," arXiv:1409.0387.

[31] G. P. Agrawal, Nonlinear Fiber Optics (Academic Press, California, 1995).

[32] T. Horiguchi, T. Kurashima, and M. Tateda, "Tensile strain dependence of Brillouin frequency shift in silica optical fibers," IEEE Photon. Technol. Lett. **1**, 107–108 (1989).

[33] T. Kurashima, T. Horiguchi, and M. Tateda, "Thermal effects on the Brillouin frequency shift in jacketed optical silica

fibers," Appl. Opt. **29**, 2219–2222 (1990).
[34] Y. Mizuno and K. Nakamura, "Core alignment of butt-coupling between single-mode and multi-mode optical fibers by monitoring Brillouin scattering signal," J. Lightwave Technol. **29**, 2616-2620 (2011).